\documentclass[sigconf,natbib=true]{acmart}

\AtBeginDocument{%
  }

\begin{document}

\title{The 2$^{nd}$ EReL@MIR Workshop on Efficient Representation Learning for Multimodal Information Retrieval}

\author{Junchen Fu}
\affiliation{%
  \institution{University of Glasgow}
  \city{Glasgow}
  \country{UK}
}
\email{j.fu.3@research.gla.ac.uk}

\author{Xuri Ge}

\affiliation{%
  \institution{Shandong University}
  \city{Shandong}
  \country{China}
  }
  
\email{xuri.ge@sdu.edu.cn}

\author{Xin Xin}
\affiliation{%
  \institution{Shandong University}
  \city{Shandong}
  \country{China}}
\email{xinxin@sdu.edu.cn}

\author{Alexandros Karatzoglou}
\affiliation{
\institution{Amazon}\streetaddress{}\city{Barcelona}\country{Spain}}
\email{alexandros.karatzoglou@gmail.com}

\author{Ioannis Arapakis}
\affiliation{
\institution{Telef\'{o}nica Scientific Research}\streetaddress{}\city{Barcelona}\country{Spain}}
\email{arapakis.ioannis@gmail.com}

\author{Xi Wang}
\affiliation{%
  \institution{University of Sheffield
}
  \city{Sheffield}
  \country{United Kingdom}}
\email{	xi.wang@sheffield.ac.uk}

\author{Qijiong Liu}
\affiliation{%
  \institution{Hong Kong Polytechnic University
}
  \city{Hong Kong}
  \country{China}}
\email{liu@qijiong.work}

\author{Qian Li}
\affiliation{
\institution{Beijing University of Posts and Telecommunications}\streetaddress{}\city{Beijing}\country{China}}
\email{li.qian@bupt.edu.cn}

\author{Joemon M. Jose}
\affiliation{%
  \institution{University of Glasgow}
  \city{Glasgow}
  \country{United Kingdom}}
\email{joemon.jose@glasgow.ac.uk}

\renewcommand{\shortauthors}{Junchen et al.}

\begin{abstract}
Multimodal representation learning has attracted increasing attention in AI, driven by the strong performance of large, pretrained multimodal foundation models such as Qwen, LLaVA, and CLIP. These models deliver impressive performance on a range of multimodal information retrieval (MIR) tasks, including web search, cross-modal retrieval, and recommender systems. Yet their massive parameter counts create major efficiency bottlenecks when adapting their representations for IR tasks during training, deployment, and inference. These limitations hinder the practical use of foundation models for representation learning in information retrieval. To address these issues, we propose organizing the EReL@MIR workshop at MM 2026, bringing together researchers from academia and industry to discuss emerging solutions, open challenges, and new efficiency metrics and benchmarks for multimodal IR representation learning in the foundation-model era. The workshop’s official website is available at \url{https://erel-mir.github.io/}.
\end{abstract}




\maketitle

\section{MOTIVATION AND PURPOSE}
Multimodal information retrieval (MIR) underpins modern multimedia services in both industry and academia, from large-scale recommendation and e-commerce search to general-purpose web and conversational search systems~\cite{covington2016deep,wang2011jigsaw,ramesh2022hierarchical,wu2021mm}. In parallel, the rapid progress of large pre-trained foundation models for language and vision--language learning (e.g., Qwen, LLaVA, and CLIP.) has significantly reshaped how multimodal representations are learned and transferred across tasks \cite{xu2025qwen25omni,liu2024visual,radford2021learning,he2025double}. These advances have enabled strong performance in MIR settings such as web search, cross-modal retrieval, and multimodal recommender systems \cite{wei2024uniir,wang2024unified,yuan2023go,wei2019mmgcn,ge2026mcot,ge20243shnet}.

Despite these gains, deploying large multimodal models in real-world MIR pipelines exposes a persistent gap between effectiveness and system feasibility. Production MIR systems must satisfy strict latency/throughput targets while controlling VRAM, storage, and end-to-end serving costs, yet efficiency is still under-reported and inconsistently evaluated in much of the literature \cite{li2024surveygen,zhu2025large}. This limits fair comparison, reproducibility, and practical adoption at scale, motivating research on efficient adaptation, fusion, compression, indexing, and scalable serving for MIR \cite{fu2025efficient,liu2024rec,yuan2024asking,fu2024iisan,fu2025crossan,zhuang2025frequency,zhuang2025bridging}.

Besides this, two emerging shifts over the past year have further expanded the MIR design space and sharpened the urgency of efficiency research. First, omni-modal models unify a broader set of modalities (text, image, audio, video) within a single framework, enabling richer interactions but exacerbating computational and serving constraints: their larger scale and prompt-dependent representations make embedding precomputation and caching difficult. \cite{xu2025qwen25omni} Second, generative MIR approaches that generate identifiers or structured candidates introduce new efficiency and reliability challenges, including decoding-time latency, output validity and controllability, and tokenization complexity. \cite{zhaimultimodal,tay2022transformer,fu2026differentiable}

We therefore propose the \textbf{2nd EReL@MIR workshop at ACM Multimedia 2026 (MM'26)} as a timely venue to advance efficiency-aware multimodal representation learning and multimedia retrieval within the MM community. The workshop complements the main MM 2026 program by strengthening core multimedia themes, including multimodal learning, vision--language modeling, multimedia retrieval, recommendation, and multimodal generation, through the lens of efficiency, deployability, and cost-aware system design \cite{fu2025efficient}. It will bring together researchers from academia and industry to address practical and theoretical challenges in building scalable multimodal foundation models and retrieval systems. The workshop has four main goals: (i) promote research on efficiency-aware multimodal learning for large-scale multimedia applications; (ii) encourage unified metrics and benchmarks that jointly evaluate effectiveness and system cost; (iii) advance deployable solutions for real-world multimedia retrieval and generation systems; and (iv) stimulate discussion on emerging omni-modal and generative paradigms under realistic computational constraints.

\vspace{-0.16in}

\section{TOPICS AND THEMES}
Building on the first edition, the 2nd EReL@MIR workshop continues to target the core challenges of efficient multimodal representation learning for MIR, spanning training, adaptation, deployment, and inference in real-world systems. This edition additionally highlights two new directions: efficient adaptation of omni-modal models (text/image/audio/video) and efficient multimodal generative MIR, where retrieval is produced via generation (e.g., identifiers or structured candidates) rather than only ranking.

\noindent \textbf{1. Efficient Multimodal Representation Adaptation based
on Multmodal Foundation Models.}
Adapting large multimodal foundation models to IR tasks is computationally expensive due to cross-modal representation learning. A key challenge is reducing adaptation cost without degrading representation quality, making parameter-efficient fine-tuning and model compression essential for scalable real-world deployment.

\noindent \textbf{2. Data-Efficiency in Multimodal Representation Learning.}
Training multimodal foundation models on large, diverse datasets is resource-intensive. Efficient data utilization, such as strategic sampling, distillation, and denoising, can reduce required data and training cost while preserving representation quality.

\noindent \textbf{3. Efficient Multimodal Fusion for Representation Learning.}
Integrating modalities like text, images, video, and audio into unified representations is computationally expensive. More efficient fusion mechanisms, such as lightweight or adaptive fusion, are needed to reduce cost while maintaining representation quality for practical IR use.

\noindent \textbf{4. Efficient Cross-Modality Interaction for MIR.} Effective cross-modal interactions are crucial but often rely on costly attention, GNNs, or contrastive learning. The goal is to design lightweight interaction mechanisms that cut training cost while preserving strong information exchange across modalities.

\noindent \textbf{5.  Real-Time Inference for Multimodal Representations.}
Real-time IR systems need fast multimodal inference, but large models often become latency bottlenecks. Pruning, quantization, and efficient encoders such as vector quantization can speed inference for deployment in latency-critical settings.

\noindent \textbf{6. Efficient MIR Foundation Models.}
Most representation learning foundation models use transformers, which are effective but computationally heavy and large. MIR needs more efficient alternatives that reduce cost without sacrificing performance.

\noindent \textbf{7. Benchmarks and Metrics for Multimodal Representation
Learning Efficiency.}
MIR lacks standardized benchmarks and comprehensive metrics for efficiency, which spans VRAM, training and inference time, and parameter count. Establishing shared benchmarks that measure both effectiveness and efficiency is essential for fair comparison and future progress.

\noindent \textbf{8. Efficient Omni-modal MIR.}
Efficiency challenges in applying omni-modal foundation models (e.g., Qwen-Omni) to IR, including compute-aware modality routing, efficient audio/video representation and fusion, and deployment under strict latency, throughput, VRAM, and serving-cost budgets (e.g., caching, quantization, and streaming/early-exit inference).

\noindent \textbf{9. Efficient Multimodal Generative MIR.}
Feasible generative retrieval for multimodal IR under tight latency and cost constraints. Topics include efficient and constrained decoding, semantic-ID and candidate generation schemes that shrink the search and generation space, compression/distillation of generator and reranker stacks for deployment, and evaluation protocols that report both quality and system cost (latency, throughput, VRAM, and serving cost) for fair comparison.

\vspace{-0.1in}
\subsection{Target Audience and Accepted Submissions}

This workshop targets researchers and practitioners in the multimedia and multimodal AI community, including those working on multimodal representation learning, vision-language models, efficient foundation models, multimedia retrieval and recommendation, and large-scale system optimization. We focus on efficiency-aware design of multimodal systems, spanning compression, parameter-efficient adaptation, indexing and serving, real-time inference, and cost-aware deployment under constraints such as latency, memory, and energy. We expect participation from both academia and industry, particularly from groups developing deployable multimodal retrieval and generative systems.

The first EReL@MIR workshop, co-located with WWW, received 9 submissions and accepted 8 papers. We attribute the limited submission volume partly to the relatively lower visibility of multimodal retrieval within the broader WebConf community. In contrast, ACM Multimedia hosts a larger and more multimedia-centric audience with strong engagement in multimodal learning and retrieval, and its community size is significantly greater than that of WWW in this area. For MM 2026, we therefore expect at least a doubling of submissions, with over 20 papers anticipated. We aim to attract contributions in efficient omni-modal modeling and multimodal generative retrieval, while continuing to welcome work on core themes of efficiency-aware multimodal representation learning. We expect to accept over 10 high-quality papers through a rigorous review process.

\vspace{-0.1in}

\subsection{Review Process}
Submissions will undergo a standard peer-review process. Each paper will be reviewed by at least three reviewers, and decisions will be made jointly by the organizers and the program committee. Accepted papers will be selected based on originality, relevance to the workshop themes, technical soundness, clarity, and potential impact. We welcome different submission types (e.g., research papers, short/position papers, and system papers) to encourage both foundational and applied contributions.
\vspace{-0.1in}

\subsection{Tentative Workshop Structure}
\noindent \textbf{Paper sessions.} The workshop will feature oral paper presentations accompanied by interactive Q\&A, with an optional poster session to further encourage in-depth discussion, community feedback, and exchange of ideas.

\noindent \textbf{Panel discussion.} A moderated panel will focus on the practical challenges and emerging research opportunities in deploying efficient multimodal and generative information retrieval systems at scale in real-world environments.

\vspace{-0.1in}
\subsection{Workshop Advertisement}

To ensure wide participation, we will promote the workshop through social media and community platforms (e.g., X/Twitter, LinkedIn, WeChat, REDnote), as well as academic mailing lists and professional networks within the ACM Multimedia and broader multimedia research community. We will conduct direct outreach to researchers and labs working on multimodal learning, vision--language models, multimedia retrieval, multimodal generation, and efficient foundation models. In addition, the workshop will be disseminated through MM-related channels, prior ACM MM workshop networks, and the organizers’ institutional and industrial collaborations to engage both academic researchers and practitioners in large-scale multimedia systems.

\vspace{-0.1in}

\section{RELATED WORKSHOPS}
The proposed second EReL@MIR workshop builds on the first edition at WWW 2025~\cite{fu20251st}, which received 9 submissions and accepted 8 papers, reflecting growing interest in efficiency-aware multimodal retrieval and recommendation. In addition, we co-hosted the EReL@MIR MMDocIR \& MM-CTR Challenge with two tracks: MMDocIR (multimodal document retrieval) attracted 41 participants and 586 submissions, while MM-CTR (multimodal click-through rate prediction) drew 176 participants and 428 submissions. The scale of participation and submissions demonstrates strong demand for standardized benchmarks, leaderboards, and reproducible evaluation in efficiency-aware multimodal systems. Given this success, we plan to continue co-hosting the challenge in conjunction with the workshop at MM 2026 to further expand community engagement and benchmarking efforts.

This year, we propose to move EReL@MIR to the MM community, as we believe it provides a more suitable and focused venue for our workshop theme than WWW. Compared with the Web-technology-centric scope of WWW, the MM community is inherently multimedia-centric, with strong emphasis on multimodal learning, vision--language models, and multimedia retrieval and recommendation. This closer alignment with our core topics is expected to attract broader and deeper engagement from researchers working on efficiency-aware multimodal systems, thereby further increasing the visibility and impact of the workshop.

\vspace{-0.1in}

\section{List of Programme Committee Members}
\begin{itemize}
    \item Jiayi Ji - National University of Singapore, Singapore
    \item Mingyue Cheng - University of Science and Technology of China, China
    \item Ying Zhou - Shandong University, China
    \item Hengchang Hu - National University of Singapore, Singapore
    \item Yumeng Wang - Leiden University, The Netherlands
    \item Fuhai Chen - Fuzhou University, China
    \item Hui Li - Xiamen University, China
    \item Songpei Xu - University of Glasgow, UK
    \item Zheng Yuan - The Hong Kong Polytechnic University, Hong Kong
    \item Jiaqi Zhang - University of Queensland, Australia
    \item Abhishek Dharmaratnakar - Google, US
\end{itemize}

\vspace{-0.1in}
\section{Diversity Aspects}
This workshop is committed to diversity across gender, institutional affiliation, and nationality in its speakers, organizers, and Programme Committee. Invited speakers will span academia and industry to ensure broad perspectives. The organizing team includes both male and female members, with nine organizers from seven institutions across the UK, Spain, and China, representing both academia and industry. Programme Committee members are selected with the same diversity goals in mind.
\vspace{-0.1in}

\section{Special Requirements and Contingency Plan}
We require a single-track room with reliable A/V (projector/screen, HDMI, and microphones), stable Wi-Fi/power, and preferably hybrid capability, as these are essential for keynote(s), paper talks, and an interactive panel/Q\&A. If any requirement is unmet, we will use backup devices/adapters and pre-collected slides, relay Q\&A via the chair if needed, replace remote segments with onsite discussion, and ensure continuity with at least two organizers onsite who can assume chairing/operations at short notice.

\section{OGANIZERS INFORMATION and ATTENDANCE LIKELIHOOD}
\noindent \textbf{Prof. Joemon M. Jose (99\%)} is a Professor at the School of Computing Science, University of Glasgow, Scotland, and a member of the Information Retrieval group. He is a Fellow of the BCS and IET professional bodies and a chartered Information Technology Professional. He has published over 300 papers with more than 11,000 Google Scholar citations, and an h-index of 54. He leads the GAIR Lab, which focuses on generative solutions for IR problems. He has been serving as the program committee chair and member for numerous top international conferences (e.g.,  MM, SIGIR, WWW, and ECIR). He also serves as a PC chair for SIGIR-AP 2024.

\noindent \textbf{Dr. Alexanderos Karatzgolou (60\%)} is a Principal Applied Scientist at Amazon. His contributions, including GRU4Rec and kernlab, have had a lasting impact on machine learning and recommender systems research. He has received over 20,000 Google Scholar citations (h-index 44). Previously, he was a Staff Research Scientist at Google DeepMind and Director of Telefónica Research in Barcelona.

\noindent \textbf{Dr. Ioannis Arapakis (60\%)} heads Telefónica Scientific Research at Telefónica Innovación Digital, leading teams in machine learning, neuroscience, cybersecurity, sustainable AI and networks. An ELLIS member and PI of the EIC Pathfinder SYMBIOTIK project on neuro-adaptive InfoVis, he previously worked at Yahoo Labs and holds an MSc (KTH) and PhD (Glasgow). He has organized the EReL@MIR workshop in WWW.

\noindent \textbf{Dr. Xin Xin (80\%)} is a Tenured Associate Professor at the School of Computer Science and Technology of Shandong University. Before that, he earned his PhD from the University of Glasgow. His current research interests include information retrieval, natural language processing, and reinforcement learning. He has published more than 40 papers in top conferences (e.g., WWW, SIGIR, ACL, WSDM) and journals (e.g., TOIS, TKDE), and received the Best Paper Honor Mention at WSDM 2024. He has organized the DRL4IR workshop in SIGIR, KEIR workshop in ECIR, and R$^{3}$AG workshop in SIGIR-AP.

\noindent \textbf{Dr. Qijiong Liu (90\%)} is currently a postdoctoral fellow at The Hong Kong Polytechnic University, where he also received his Ph.D. degree. He earned his B.Eng. and M.Eng. degrees from Zhejiang University. His research interests lie in recommender systems and generative retrieval. He has contributed to the field through his work published in prestigious conferences such as WWW, WSDM, AAAI, and IJCAI. He also co-organized a tutorial on multimodal recommendation at KDD 2024.

\noindent \textbf{Dr. Qian Li (99\%)} is currently an Assistant Professor at Beijing University of Posts and Telecommunications (BUPT). Her research interests include large language models, multimodal learning, and multimodal information retrieval. She has published 50+ papers in leading conferences and journals, including NeurIPS, ACM MM, ACL, WWW,  AAAI, and IEEE Transactions venues. She has served as an Area Chair and SPC member for top-tier conferences and journals, such as ACL, NeurIPS, AAAI, IJCAI, and EMNLP. She received a Best Paper Nomination at CIKM 2022.

\noindent \textbf{Dr. Xi Wang (99\%)} is a Lecturer at the University of Sheffield. He obtained his PhD from the University of Glasgow. He was a postdoctoral researcher at University College London, working on task-oriented dialogue systems with Prof. Emine Yilmaz. His research interests include Interactive NLP, Conversational AI, and personalization. His work has been published in top-tier venues such as SIGIR, AAAI, WWW, CIKM, and ECIR, etc.

\noindent \textbf{Dr. Xuri Ge (99\%)} is now a tenure-track assistant professor at Shandong University. He earned his PhD at the University of Glasgow and received an M.S. degree from Xiamen University. His current research interests include computer vision, multimodal representation learning, and information retrieval. He has contributed to several leading conferences and journals, including NeurIPS, ACM MM, SIGIR,  CIKM, and IP\&M, etc. He also serves as the PC member and reviewer for tier-1 conferences and journals, such as NeurIPS, WWW, MM, AAAI, ICLR, TKDE, TOIS, etc. He has organized the NIP@IR workshop at SIGIR, 3DMM Workshop at ICME, EReL@MIR workshop at WWW, and R$^{3}$AG workshop at SIGIR-AP.

\noindent \textbf{Junchen Fu (99\%)} is a final-year PhD candidate supervised by Prof.~Joemon Jose at the University of Glasgow. He was a visiting scholar at Leiden University, advised by Prof.~Zhaochun Ren and Prof.~Suzan Verberne. He has published at leading AI venues including ICML, MM, SIGIR, WWW, and TKDE, etc, and serves as a PC member and reviewer for top-tier venues such as MM, SIGIR, WWW, WSDM, TKDE, and TOIS, etc. He organized the EReL@MIR workshop at WWW and the R$^{3}$AG workshop at SIGIR-AP.

\vspace{-0.1in}
\bibliographystyle{ACM-Reference-Format}
\bibliography{sample-base}

@article{ge20243shnet,
  title={3SHNet: Boosting image--sentence retrieval via visual semantic--spatial self-highlighting},
  author={Ge, Xuri and Xu, Songpei and Chen, Fuhai and Wang, Jie and Wang, Guoxin and An, Shan and Jose, Joemon M},
  journal={Information Processing \& Management},
  volume={61},
  number={4},
  pages={103716},
  year={2024},
  publisher={Elsevier}
}

@inproceedings{ge2026mcot,
  title={MCoT-MVS: Multi-level Vision Selection by Multi-modal Chain-of-Thought Reasoning for Composed Image Retrieval},
  author={Ge, Xuri and Wang, Chunhao and Wang, Xindi and Qin, Zheyun and Chen, Zhumin and Xin, Xin},
  booktitle={Proceedings of the ACM Web Conference 2026},
  pages={2105--2113},
  year={2026}
}

@inproceedings{zhuang2025bridging,
  title={Bridging the Gap: Teacher-Assisted Wasserstein Knowledge Distillation for Efficient Multi-Modal Recommendation},
  author={Zhuang, Ziyi and Du, Hanwen and Han, Hui and Li, Youhua and Fu, Junchen and Jose, Joemon M and Ni, Yongxin},
  booktitle={Proceedings of the ACM on Web Conference 2025},
  pages={2464--2475},
  year={2025}
}

@inproceedings{zhuang2025frequency,
  title={Frequency-Decoupled distillation for efficient multimodal recommendation},
  author={Zhuang, Ziyi and Li, Hongji and Fu, Junchen and Liu, Jiacheng and Jose, Joemon M and Li, Youhua and Ni, Yongxin},
  booktitle={Proceedings of the 34th ACM International Conference on Information and Knowledge Management},
  pages={4571--4581},
  year={2025}
}

@inproceedings{fu20251st,
  title={The 1st erel@ mir workshop on efficient representation learning for multimodal information retrieval},
  author={Fu, Junchen and Ge, Xuri and Xin, Xin and Yu, Haitao and Feng, Yue and Karatzoglou, Alexandros and Arapakis, Ioannis and Jose, Joemon},
  booktitle={Companion Proceedings of the ACM on Web Conference 2025},
  pages={2149--2152},
  year={2025}
}

@article{fu2025crossan,
  title={Crossan: Towards efficient and effective adaptation of multiple multimodal foundation models for sequential recommendation},
  author={Fu, Junchen and Ni, Yongxin and Jose, Joemon M and Arapakis, Ioannis and Zheng, Kaiwen and Li, Youhua and Ge, Xuri},
  journal={arXiv preprint arXiv:2504.10307
        
        
        
        
        
        
        
        },
  year={2025}
}

@inproceedings{he2025double,
  title={Double-filter: Efficient fine-tuning of pre-trained vision-language models via patch\&layer filtering},
  author={He, Yaoqin and Fu, Junchen and Zheng, Kaiwen and Xu, Songpei and Chen, Fuhai and Li, Jie and Jose, Joemon M and Ge, Xuri},
  booktitle={Forty-second International Conference on Machine Learning},
  year={2025}
}

@article{fu2026differentiable,
  title={Differentiable Semantic ID for Generative Recommendation},
  author={Fu, Junchen and Ge, Xuri and Karatzoglou, Alexandros and Arapakis, Ioannis and Verberne, Suzan and Jose, Joemon M and Ren, Zhaochun},
  journal={arXiv preprint arXiv:2601.19711
        
        
        
        
        
        },
  year={2026}
}

@inproceedings{zhaimultimodal,
  title={Multimodal Quantitative Language for Generative Recommendation},
  author={Zhai, Jianyang and Mai, Zi-Feng and Wang, Chang-Dong and Yang, Feidiao and Zheng, Xiawu and Li, Hui and Tian, Yonghong},
  booktitle={The Thirteenth International Conference on Learning Representations}
}

@inproceedings{wang2011jigsaw,
  title={JIGSAW: interactive mobile visual search with multimodal queries},
  author={Wang, Yang and Mei, Tao and Wang, Jingdong and Li, Houqiang and Li, Shipeng},
  booktitle={Proceedings of the 19th ACM international conference on Multimedia},
  pages={73--82},
  year={2011}
}

@article{fu2025efficient,
  title={Efficient and effective adaptation of multimodal foundation models in sequential recommendation},
  author={Fu, Junchen and Ge, Xuri and Xin, Xin and Karatzoglou, Alexandros and Arapakis, Ioannis and Zheng, Kaiwen and Ni, Yongxin and Joemon, Joemon M Jose},
  journal={IEEE Transactions on Knowledge and Data Engineering},
  year={2025},
  publisher={IEEE}
}

@inproceedings{yuan2024asking,
  title={Asking Multimodal Clarifying Questions in Mixed-Initiative Conversational Search},
  author={Yuan, Yifei and Siro, Clemencia and Aliannejadi, Mohammad and Rijke, Maarten de and Lam, Wai},
  booktitle={Proceedings of the ACM on Web Conference 2024},
  pages={1474--1485},
  year={2024}
}

@article{zhu2025large,
  title={Large language models for information retrieval: A survey},
  author={Zhu, Yutao and Yuan, Huaying and Wang, Shuting and Liu, Jiongnan and Liu, Wenhan and Deng, Chenlong and Chen, Haonan and Liu, Zheng and Dou, Zhicheng and Wen, Ji-Rong},
  journal={ACM Transactions on Information Systems},
  volume={44},
  number={1},
  pages={1--54},
  year={2025},
  publisher={ACM New York, NY}
}

@article{liu2024visual,
  title={Visual instruction tuning},
  author={Liu, Haotian and Li, Chunyuan and Wu, Qingyang and Lee, Yong Jae},
  journal={Advances in neural information processing systems},
  volume={36},
  year={2024}
}

@inproceedings{wei2024uniir,
  title={Uniir: Training and benchmarking universal multimodal information retrievers},
  author={Wei, Cong and Chen, Yang and Chen, Haonan and Hu, Hexiang and Zhang, Ge and Fu, Jie and Ritter, Alan and Chen, Wenhu},
  booktitle={European Conference on Computer Vision},
  pages={387--404},
  year={2024},
  organization={Springer}
}

@inproceedings{wang2024unified,
  title={Unified embeddings for multimodal retrieval via frozen LLMs},
  author={Wang, Ziyang and Elfardy, Heba and Dreyer, Markus and Small, Kevin and Bansal, Mohit},
  booktitle={Findings of the Association for Computational Linguistics: EACL 2024},
  pages={1537--1547},
  year={2024}
}

@inproceedings{covington2016deep,
  title={Deep neural networks for youtube recommendations},
  author={Covington, Paul and Adams, Jay and Sargin, Emre},
  booktitle={Proceedings of the 10th ACM conference on recommender systems},
  pages={191--198},
  year={2016}
}

@article{liu2024rec,
  title={Rec-GPT4V: Multimodal Recommendation with Large Vision-Language Models},
  author={Liu, Yuqing and Wang, Yu and Sun, Lichao and Yu, Philip S},
  journal={arXiv preprint arXiv:2402.08670
        
        
        
        
        
        
        
        
        
        },
  year={2024}
}

@inproceedings{fu2024iisan,
  title={IISAN: Efficiently Adapting Multimodal Representation for Sequential Recommendation with Decoupled PEFT},
  author={Fu, Junchen and Ge, Xuri and Xin, Xin and Karatzoglou, Alexandros and Arapakis, Ioannis and Wang, Jie and Jose, Joemon M},
  booktitle={Proceedings of the 47th International ACM SIGIR Conference on Research and Development in Information Retrieval},
  pages={687--697},
  year={2024}
}

@article{ramesh2022hierarchical,
  title={Hierarchical text-conditional image generation with clip latents},
  author={Ramesh, Aditya and Dhariwal, Prafulla and Nichol, Alex and Chu, Casey and Chen, Mark},
  journal={arXiv preprint arXiv:2204.06125},
  volume={1},
  number={2},
  pages={3},
  year={2022}
}

@inproceedings{radford2021learning,
  title={Learning transferable visual models from natural language supervision},
  author={Radford, Alec and Kim, Jong Wook and others},
  booktitle={International conference on machine learning},
  pages={8748--8763},
  year={2021},
  organization={PMLR}
}

@article{wu2021mm,
  title={Mm-rec: multimodal news recommendation},
  author={Wu, Chuhan and Wu, Fangzhao and Qi, Tao and Huang, Yongfeng},
  journal={arXiv preprint arXiv:2104.07407},
  year={2021}
}

@inproceedings{wei2019mmgcn,
  title={MMGCN: Multi-modal graph convolution network for personalized recommendation of micro-video},
  author={Wei, Yinwei and Wang, Xiang and Nie, Liqiang and He, Xiangnan and Hong, Richang and Chua, Tat-Seng},
  booktitle={Proceedings of the 27th ACM International Conference on Multimedia},
  pages={1437--1445},
  year={2019}
}

@String{Computer = "{IEEE} Computer" }

@String{Springer = "Springer-Verlag" }

@inproceedings{yuan2023go,
  title={Where to go next for recommender systems? id-vs. modality-based recommender models revisited},
  author={Yuan, Zheng and Yuan, Fajie and Song, Yu and Li, Youhua and Fu, Junchen and Yang, Fei and Pan, Yunzhu and Ni, Yongxin},
  booktitle = {Proceedings of the 46th International ACM SIGIR Conference on Research and Development in Information Retrieval},
 pages = {2639–2649},
  year={2023}
}

@article{xu2025qwen25omni,
  title        = {Qwen2.5-Omni Technical Report},
  author       = {Xu, Jin and Guo, Zhifang and others},
  journal      = {arXiv preprint arXiv:2503.20215
        
        
        
        
        
        
        
        },
  year         = {2025}
}

@article{tay2022transformer,
  title={Transformer memory as a differentiable search index},
  author={Tay, Yi and Tran, Vinh and Dehghani, Mostafa and Ni, Jianmo and Bahri, Dara and Mehta, Harsh and Qin, Zhen and Hui, Kai and Zhao, Zhe and Gupta, Jai and others},
  journal={Advances in Neural Information Processing Systems},
  volume={35},
  pages={21831--21843},
  year={2022}
}

@article{li2024surveygen,
  title        = {A Survey of Generative Search and Recommendation in the Era of Large Language Models},
  author       = {Li, Yongqi and Lin, Xinyu and Wang, Wenjie and Feng, Fuli and Pang, Liang and Li, Wenjie and Nie, Liqiang and He, Xiangnan and Chua, Tat-Seng},
  journal      = {arXiv preprint arXiv:2404.16924
        
        
        
        
        
        },
  year         = {2024}
}

\end{document}